\documentclass[aps,pre,reprint,longbibliography]{revtex4-2}

\usepackage{graphicx}\graphicspath{ {figures/} }
\usepackage{hyperref}
\usepackage{amsmath}
\usepackage{amssymb}
\hypersetup{colorlinks,allcolors=blue,breaklinks}

\newcommand{\be}{\begin{equation}}
\newcommand{\ee}{\end{equation}}
\newcommand{\ba}{\begin{align}}
\newcommand{\ea}{\end{align}}
\newcommand{\bi}{\begin{itemize}}
\newcommand{\ei}{\end{itemize}}

\newcommand{\bla}{bla\\bla\\bla\\bla\\bla}

\begin{document}

\title{Linear optimal protocol for physical constraints in weakly driven processes}

\author{Pierre Naz\'e}
\email{pnaze@ufpa.br}

\affiliation{\it Universidade Federal do Par\'a, ICEN, Faculdade de F\'isica,
Av. Augusto Correa, 1, Guam\'a, 66075-110, Bel\'em, Par\'a, Brazil}

\date{\today}

\begin{abstract}
The minimization of irreversible work in weakly driven systems within linear response under physical constraints on the protocol derivative is studied. The problem reduces to a shifted eigenvalue equation involving the relaxation function. Owing to its dependence on time differences and its evenness, the relaxation kernel is naturally defined over a symmetric interval, where a periodic representation arises as a consistent closure that restores continuous translational invariance. Also, it shows how the irreversible work is defined in practice. Within this framework, the operator becomes diagonal in a Fourier basis. The global optimal solution is the zero mode, yielding a constant driving speed and a linear protocol. The corresponding optimal work depends only on the integrated relaxation function. Numerical results obtained via genetic programming confirm the robustness of this solution across different kernels.
\end{abstract}

\maketitle

\section{Introduction}

The control of thermodynamic processes at small scales has become a central topic in nonequilibrium physics, where fluctuations and dissipation play a dominant role \cite{Deffner2020, Blaber2023, Alvarado}. A fundamental question in this context is how to design external protocols that drive a system between two states while minimizing energetic costs. Within the regime of weak driving, linear response theory provides a general and powerful framework to address this problem, expressing the irreversible work in terms of equilibrium correlation functions \cite{naze2020}.

A number of studies have shown that optimal protocols may exhibit nontrivial features, such as sharp variations near the boundaries of the process \cite{Schmiedl2007, Gomez-Marin2008, Naze2024, Naze2025a}. While such behaviors can reduce dissipation, they are often not physically realizable, as they may require arbitrarily large driving rates. This motivates the introduction of constraints that reflect realistic limitations on the control parameters~\cite{MuratoreGinanneschi2026}.

In this work, we consider a constrained optimization problem in which the derivative of the protocol is bounded through a quadratic condition, while the total variation of the control parameter is fixed. Within linear response, this leads to a variational problem for the irreversible work functional, whose stationarity condition takes the form of a shifted eigenvalue equation involving the relaxation function, which encodes the memory of the system.

A central aspect of this formulation concerns the structure of the relaxation kernel in finite time processes. The irreversible work depends on the function $\Psi_0(t-u)$, which is determined by equilibrium correlations. This dependence on time differences implies that, even though the protocol is defined over the interval $[0,\tau]$, the kernel is naturally defined over the symmetric domain $[-\tau,\tau]$. In addition, the relaxation function is even, reflecting the symmetry of equilibrium correlations. These two properties together define the natural domain and symmetry of the operator associated with the variational problem.

When the dynamics is restricted to a finite interval, however, boundary effects break the translational invariance that is implicit in the dependence on time differences. The appropriate way to restore this structure is not to impose an external condition, but to consistently complete the kernel on its natural domain. This is achieved by extending the relaxation function periodically over $[-\tau,\tau]$. Such a construction preserves both the dependence on time differences and the evenness of the kernel, while ensuring continuity at the boundaries and restoring translational invariance.

From this perspective, the periodic representation of the relaxation function should not be viewed as a mathematical artifact, but as a direct consequence of its intrinsic structure. Once this representation is adopted, the integral operator becomes a convolution operator on a periodic domain, and the Fourier basis emerges as its natural spectral representation. The diagonalization of the problem then follows directly from these structural properties.

From an operational perspective, the periodic representation of the relaxation function can be understood as a natural consequence of how irreversible work is defined in practice. The averaged work is obtained as an ensemble average over many realizations of the same finite-time protocol, each performed over the interval $[0,\tau]$ and starting from an identical initial statistical state. After each realization, the system is effectively reset, and the protocol is applied again.

In this sense, the ensemble of realizations introduces a block structure in time, where each block corresponds to a single execution of the protocol. Since the relaxation function depends only on time differences, and all realizations are statistically equivalent, there is no distinguished origin of time across the ensemble. As a result, time differences are represented on a periodic domain, leading to an effective periodic representation of the kernel.

Within this framework, the optimization problem can be analyzed in terms of the eigenmodes of the relaxation kernel. The positivity of the kernel plays a decisive role, as it selects the zero mode as the unique minimizer under the imposed constraints. As a consequence, the optimal protocol corresponds to a constant driving speed, leading to a linear control parameter in time. This result is not imposed, but follows from the symmetry and spectral properties of the operator. Similar results have already been found in other contexts, such as slowly varying limit using geometric framework approach~\cite{Sivak2012}, and Markovian systems with low-noise limit~\cite{Monter2026a}. 

While linear protocols are known to arise in linear-response frameworks under specific conditions, such as constant thermodynamic metric or slow driving, in those cases linearity follows from the locality of the functional or geometric arguments. In contrast, here linearity emerges from the spectral structure of a nonlocal quadratic functional, where the positivity of the relaxation kernel selects the zero mode as the global minimizer under constraints.

To assess the generality of these findings, we complement the analytical treatment with numerical optimization based on genetic programming \cite{Koza1994}. This approach allows us to explore a broad class of admissible protocols and relaxation functions, including kernels with distinct temporal structures. In all cases examined, the optimal protocol converges to a linear form, and the corresponding work agrees with the analytical prediction with high precision, indicating that the result is robust and not dependent on specific features of the kernel.

This paper is organized as follows. We first introduce the irreversible work functional and discuss the structural properties of the relaxation function in finite time processes. We then formulate the constrained variational problem and derive the associated Euler-Lagrange equation. Next, we solve the resulting shifted eigenvalue problem using a spectral decomposition. Finally, we present numerical results that support the analytical findings and discuss their implications.

\section{Preliminaries}

We consider a system driven by an external control parameter $\lambda(t)$ over a finite time interval $t \in [0,\tau]$. In the regime of weak driving, linear response theory provides an expression for the irreversible work in terms of equilibrium fluctuations. Defining the driving speed as $v(t) \equiv \dot{\lambda}(t)$, the irreversible work functional can be written as
\begin{equation}
W_{\mathrm{irr}}[v] = \frac{1}{2} \int_0^\tau \int_0^\tau \Psi_0(t - u)\, v(t)\, v(u)\, du\, dt.
\end{equation}

The relaxation function $\Psi_0(t)$ encodes the temporal correlations of the generalized force conjugate to $\lambda$. It is assumed to be even and positive definite, as required by equilibrium linear response theory~\cite{naze2020}. If the system starts from the thermal equilibrium, this formula for the irreversible work holds for the two-point measurement scheme of quantum work~\cite{guarnieri2024generalised}. Also, the system can be open or thermally isolated.

A key point in the present formulation is the structure induced by the dependence of the kernel on time differences. Since the work functional involves $\Psi_0(t - u)$ with $t,u \in [0,\tau]$, the argument of the relaxation function naturally spans the symmetric interval $t - u \in [-\tau,\tau]$. Therefore, the kernel is intrinsically defined over this symmetric domain.

In addition, the evenness of the relaxation function,
\begin{equation}
\Psi_0(t) = \Psi_0(-t),
\end{equation}
ensures that it is continuous at the boundaries of the interval $[-\tau,\tau]$. These two properties together define the natural space on which the integral operator acts.

However, when restricted to a finite interval, the associated operator does not possess full translational invariance due to boundary effects. To restore this structure without altering the intrinsic properties of the kernel, we consider the minimal extension that preserves both the dependence on time differences and the evenness of the relaxation function. This is achieved through the periodic closure
\begin{equation}
\Psi_0(t) = \Psi_0(t + 2n\tau),
\end{equation}
with $t \in [-\tau,\tau]$ and $n \in \mathbb{Z}$.

This construction should not be interpreted as an additional assumption, but rather as a consistent completion of the kernel on its natural domain. It preserves the local structure of correlations while restoring translational invariance of the operator. As a consequence, the integral operator acquires a convolution structure on a periodic domain, which allows for a spectral decomposition in a Fourier basis.

From an operational standpoint, this periodic representation also reflects how the irreversible work is constructed in practice. The work is defined as an ensemble average over many realizations of the same finite-time protocol, each carried out over the interval $[0,\tau]$ and starting from an identical initial state. After each realization, the system is effectively reset and the protocol is applied again, leading to a sequence of statistically equivalent blocks in time. Since the relaxation function depends only on time differences, and no particular realization is distinguished within the ensemble, time differences are naturally represented on a periodic domain. In this sense, the periodic extension does not introduce new physical assumptions, but rather provides a representation that is consistent with the statistical structure of repeated measurements while preserving the intrinsic temporal correlations within each realization.

Throughout this work, we adopt this representation as the natural framework to describe the interplay between memory and finite time driving in weakly driven systems.

\section{Euler-Lagrange equation}

We start considering the linear-response functional for the irreversible work
\begin{equation}
W_{\mathrm{irr}}[v]
=
\frac12\int_0^\tau \int_0^\tau \Psi_0(t-u)v(u)v(t)dudt,
\end{equation}
where the relaxation function $\Psi_0(t)$ is assumed to be even, positive-definite~\cite{naze2020}, and periodic with period $2\tau$. The derivative of the protocol $v(t)$ is subject to the constraints
\begin{equation}
\int_0^\tau v(t)dt = \delta\lambda,    
\end{equation}
which guarantees that the protocol is fixed at the borders. Also
\begin{equation}
\int_0^\tau v(t)^2dt = C,
\end{equation}
where $C<\infty$, which ensures that the optimal protocol does not exhibit divergent derivatives. These are problematic issues that often appear in the literature~\cite{Gomez-Marin2008,Naze2024,Naze2025a}.

We introduce a Lagrange multipliers $\alpha_1/2$ and $\alpha_2/2$ and define the auxiliary functional
\be
\begin{split}
\mathcal{J}[v]
=
\frac12\int_0^\tau &\int_0^\tau \Psi_0(t-u)v(u)v(t)dudt\\
&-\frac{\alpha_1}{2}\left(\int_0^\tau v(t) dt - \delta\lambda\right)\\
&-\frac{\alpha_2}{2}\left(\int_0^\tau v(t)^2 dt - C\right).
\end{split}
\ee
The stationarity condition $\delta \mathcal{J}=0$ implies in the following Euler-Lagrange equation
\begin{equation}
\int_0^\tau \Psi_0(t-u)\,v(u)\,du = \alpha_1+\alpha_2\, v(t).
\end{equation}
The problem of finding the optimal protocol is therefore a shifted eigenvalue equation.

Although the Euler-Lagrange equation is defined for $t \in [0,\tau]$, it is convenient to embed the problem into an extended domain in order to exploit its underlying structure. This can be achieved by introducing an even extension of the protocol, $\overline{v}(t)=v(|t|)$ for $t\in[-\tau,\tau]$, together with a periodic extension of the relaxation function with period $2\tau$. With this construction, the integral operator acquires a convolution structure on the extended domain while preserving its dependence on time differences and the evenness of the kernel. Importantly, this extension is consistent in the sense that, for all $t \in [0,\tau]$, the action of the extended operator coincides exactly with the original one, so that no additional boundary contributions are introduced (see Appendix ~\ref{app:A}). As a consequence, any solution of the extended Euler-Lagrange equation, when restricted to the interval $[0,\tau]$, satisfies the original variational problem. Since the kernel is positive definite, the minimizer is unique, and therefore the solution obtained in the extended setting necessarily coincides with the solution of the original problem.

\section{Shifted Eigenvalue Problem}

To solve the shifted eigenvalue problem, since the relaxation function is even and $2\tau$-periodic, it can be expanded in a cosine series
\begin{equation}
\Psi_0(t)=\sum_{n=0}^\infty a_n \cos\left(\frac{n\pi t}{\tau}\right).
\end{equation}

Expand now the extension $\overline{v}(t)=v(|t|)$ on the same Fourier basis on $[-\tau,\tau]$
\begin{equation}
\overline{v}(t)=c_0+\sum_{n=1}^\infty c_n \cos\left(\frac{n\pi t}{\tau}\right).
\end{equation}
Using orthogonality, the operator diagonalizes. The eigenvalues are
\begin{equation}
\beta_n =
\begin{cases}
\tau a_0 & n=0,\\
\frac{\tau}{2}a_n & n\ge1,
\end{cases}
\end{equation}
with the cosine basis corresponding to the eigenfunctions associated. Matching with the right-hand side of the Euler-Lagrange equation yields
\begin{align}
\beta_0 c_0 &= \alpha_1 c_0+\alpha_2, \\
(\beta_n-\alpha_1)c_n &=0, \quad n\ge1,
\end{align}
which furnish the Lagrange multipliers $\alpha_1$ and $\alpha_2$.

\subsection{General solution}

\subsubsection{Case A: $\alpha_1 \neq \beta_n$ for all $n\ge1$}
In this case, $c_n=0$, for $n\ge1$. It remains only the constant eigenfunction
\begin{equation}
\overline{v}(t)=c_0, \qquad
c_0=\frac{\alpha_2}{\beta_0-\alpha_1}.
\end{equation}
Also, $\alpha_1$ is free, while $\alpha_2=c_0(\beta_0-\alpha_1)$, with $c_0$ to be determined by the first constraint.

\subsubsection{Case B: $\alpha_1=\beta_m$}
The solution reduces to a single mode $m$
\begin{equation}
\overline{v}(t)=c_0 + c_m\cos\left(\frac{m\pi t}{\tau}\right),\quad c_0=\frac{\alpha_2}{\beta_0-\beta_m}.
\end{equation}
Also, $\alpha_1=\beta_m$, while $\alpha_2=c_0(\beta_0-\beta_m)$, with $c_0$ to be determined by the first constraint.

\subsection{Constraints}

To determine the Lagrange multipliers, we impose:
\begin{align}
\int_0^\tau v(t)\,dt &= \delta\lambda,\\
\int_0^\tau v(t)^2\,dt &= C.
\end{align}

\subsubsection{Case A: $\alpha_1 \neq \beta_n$ for all $n\ge1$}

\paragraph{Mean constraint}
\begin{equation}
c_0=\frac{\delta\lambda}{\tau}.
\end{equation}

\paragraph{Quadratic constraint}
\begin{equation}
C = \int_0^\tau v(t)^2 dt = \frac{\delta\lambda^2}{\tau}.
\end{equation}

For this case, the number $C$ is fixed by the parameters $\delta\lambda$ and $\tau$ of the system, which are previously chosen by the agent that performs the work. In practice, it is a compatibility condition for obtaining the zero-mode solution to the shifted eigenvalue problem.

\subsubsection{Case B: $\alpha_1=\beta_m$}

\paragraph{Mean constraint}
\begin{equation}
c_0=\frac{\delta\lambda}{\tau}.
\end{equation}

\paragraph{Quadratic constraint}
\begin{equation}
C =\tau c_0^2+\frac{\tau}{2}c_m^2.
\end{equation}
Thus
\begin{equation}
C=\frac{\delta\lambda^2}{\tau}+\frac{\tau}{2}c_m^2.
\end{equation}
In this case, $c_m$ can be adjusted according to the previously chosen $C$. The eigenfunction depends now on an adjusted single mode $m$.

The quadratic constraint plays a dual role in the present formulation. On the one hand, it ensures physical admissibility by preventing divergent driving rates. On the other hand, it defines the subset of admissible functions over which the quadratic form associated with the relaxation kernel is minimized. In this sense, the constraint determines which eigenmodes of the operator can contribute to the solution.

In particular, the condition $C = \delta\lambda^2/\tau$ corresponds to the case where the constraint is exactly saturated by the constant function, so that the minimizer lies entirely in the zero mode. For larger values of $C$, additional modes may appear in stationary solutions, but they necessarily increase the value of the functional due to the positivity of the kernel. Thus, the constraint does not restrict the existence of solutions, but rather selects whether the global minimizer is realized purely as the ground state of the operator.

\section{Optimal work and protocol}

The irreversible work is
\begin{equation}
W_{\mathrm{irr}}=\frac12\int_0^\tau \int_0^\tau \Psi_0(t-u)v(u)v(t)\,du\,dt.
\end{equation}
Using the eigenmode decomposition:
\begin{equation}
W_{\mathrm{irr}}
=\frac12\beta_0 c_0^2+\frac{\tau}{4}\sum_{n\ge1}\beta_n c_n^2.
\end{equation}
For an admissible optimal solution (a single mode $m$):
\begin{equation}
W_{\mathrm{irr}}^*
=\frac12\beta_0 \frac{\delta\lambda^2}{\tau^2}
+\frac{\tau}{4}\beta_m c_m^2.
\end{equation}
Since the relaxation function is positive-definite, $\beta_0$ and $\beta_m$ are positive, and the global minimum irreversible work occurs for $c_m=0$. In other words, the global minimum value of such a quadratic form is given by the smallest eigenvalue of the positive-definite kernel, which is $\beta_0$. The eigenfunction related to this situation is the constant one
\begin{equation}
v^*(t)=\frac{\delta\lambda}{\tau}.
\end{equation}
The global optimal work can be written as
\be
W_{\mathrm{irr}}^*=\frac12{\left(\frac{\delta\lambda}{\tau}\right)}^2\int_0^\tau \int_0^\tau \Psi_0(t-u)\,du\,dt,
\ee
so it depends only on the integrated relaxation function. For the slowly varying limit, $\Psi_0(t)\sim \delta(t)$, and the irreversible work scales as $\sim \tau^{-1}$, in agreement with Sivak and Crooks work~\cite{Sivak2012}.

\section{Remarks on alternative bases}

It is instructive to contrast the Fourier representation with alternative complete bases on a finite interval, such as Chebyshev polynomials~\cite{deffner2018}. Expanding the protocol as
\be
v(t) = \sum_{n=0}^\infty d_n\, T_n\!\left(\frac{2t}{\tau}-1\right),
\ee
one still obtains a well-defined representation of the variational problem on $[0,\tau]$. However, in this basis the integral operator associated with the relaxation function is no longer diagonal, even after the periodic extension of the kernel. Instead, one obtains a full matrix representation (see Appendix~\ref{app:B}), reflecting the fact that Chebyshev polynomials are not eigenfunctions of convolution operators.

This observation highlights that the diagonal structure obtained in the Fourier basis is not generic, but rather a direct consequence of the translational invariance restored by the periodic representation. Importantly, the optimal solution remains unchanged, since it is determined by the spectral properties of the operator itself and not by the choice of basis. In particular, the constant function, corresponding to the zero mode, remains the global minimizer, although its identification is no longer immediate without diagonalization.

Also, in the absence of periodic extension, the restriction of the problem to a finite interval introduces boundary-induced artifacts in spectral representations, commonly associated with Gibbs-type oscillations and spectral leakage. In practical applications, such effects are often mitigated by the use of windowing or apodization techniques, which effectively modify the basis to suppress boundary contributions. In contrast, the periodic representation adopted here eliminates these artifacts at the operator level by restoring translational invariance.

\section{Examples}

To validate the analytical predictions, we employed genetic programming as a flexible numerical framework to explore the space of admissible protocols~\cite{pnaze_bcwd}. The objective is the minimization of the irreversible work functional under the physical constraints. In particular, we look at the global minimum solution, by imposing $C=\delta\lambda^2/\tau$, and a generic protocol $g(t)$ fixed at the borders
\be
g(t) = \frac{t}{\tau}+\frac{t}{\tau}\left(1-\frac{t}{\tau}\right)h(t),
\ee
where $h(t)$ is a well-behaved random function built by genetic programming.

Constraint satisfaction was enforced through a feasibility pressure mechanism within the evolutionary algorithm. In practice, candidate protocols that violate the constraints are penalized in their fitness, ensuring that the population evolves toward the physically admissible manifold defined by fixed total variation and bounded quadratic driving speed. This procedure does not impose any specific functional form on the protocol, but rather restricts the search to the set of admissible solutions of the constrained variational problem.

We emphasize that feasibility pressure acts only to enforce the constraints and does not bias the optimization toward linear solutions. In particular, no assumptions about smoothness, monotonicity, or spectral content are imposed. The space explored by the genetic programming includes highly nontrivial functional forms, including oscillatory and non-monotonic protocols. Despite this, the optimization consistently converges to a constant driving speed.

\begin{figure}
    \centering
    \includegraphics[scale=0.75]{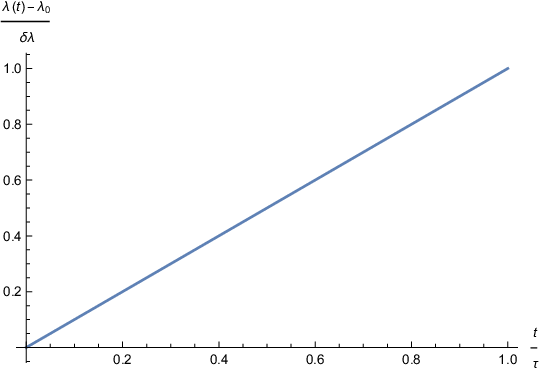}
    \caption{Linear protocol observed for all relaxation functions proposed in this work and $\tau/\tau_R=0.1,1,10$.}
    \label{fig:1}
\end{figure}

To assess the robustness of the results, we considered several classes of relaxation functions with qualitatively distinct temporal structures, including exponential decay, oscillatory behavior, and slowly decaying correlations. All of these relaxation functions are positive-definite, as a requirement to satisfy the Second Law of Thermodynamics~\cite{naze2020}. The results are summarized in Tables~\ref{table:1},~\ref{table:2},~\ref{table:3},~\ref{table:4} and illustrated in Fig.~\ref{fig:1}. In all cases, the optimal protocol converges to a linear form,
\begin{equation}
\lambda^*(t) = \lambda_0 + \frac{\delta\lambda}{\tau} t,
\end{equation}
with numerical estimates of the irreversible work in excellent agreement with the analytical prediction.

Furthermore, the constraints are satisfied with high numerical precision (typically below $10^{-6}$), confirming that the optimization effectively operates within the admissible space. The agreement between numerical and analytical results across all tested kernels provides strong evidence that the emergence of the linear protocol is a robust feature of the constrained optimization problem, rather than a consequence of specific algorithmic choices.

The excellent agreement between numerical estimates and exact analytical values provides strong evidence that the linear protocol is not only optimal but also universal within the present framework: linear response theory, bounded quadratic driving speed, fixed boundaries, and periodic kernel.

\begin{table}[h!]
\centering
\begin{tabular}{|c|c|c|c|}
\hline
\multicolumn{4}{|c|}{$\Psi_0(t)=\exp{(-|t|/\tau_R)}$}  \\
\hline
$\tau/\tau_R$ & $0.1$ & $1$ & $10$ \\
\hline
$W_{\rm irr}^*$ &0.4837$\pm$0.0001  &0.3679$\pm$0.0001  &0.0903$\pm$0.0003  \\
\hline
$W_{\mathrm{exact}}^*$ &0.483742  &0.367879  &0.0900005  \\
\hline
QC &$10^{-9}$  & $10^{-6}$  &$10^{-8}$  \\
\hline
\end{tabular}
\caption{Values of $W_{\rm irr}^*$, quadratic constraint (QC), and $W_{\mathrm{exact}}^*$ for different $\tau/\tau_R$ for $\Psi_0(t)=\exp{(-|t|/\tau_R)}$. For $\tau/\tau_R=0.1,1,10$, a straight line was observed (see Fig.~\ref{fig:1}).}
\label{table:1}
\end{table}

\begin{table}[h!]
\centering
\begin{tabular}{|c|c|c|c|}
\hline
\multicolumn{4}{|c|}{$\Psi_0(t)=J_0{(t/\tau_R)}$}  \\
\hline
$\tau/\tau_R$ & $0.1$ & $1$ & $10$ \\
\hline
$W_{\rm irr}^*$ &0.4998$\pm$0.0001  &0.4797$\pm$0.0001  &0.102395$\pm$0.000003  \\
\hline
$W_{\mathrm{exact}}^*$ &0.499792  &0.47968  &0.102354  \\
\hline
QC &$10^{-9}$  & $10^{-7}$  &$10^{-15}$  \\
\hline
\end{tabular}
\caption{Values of $W_{\rm irr}^*$, quadratic constraint (QC), and $W_{\mathrm{exact}}^*$ for different $\tau/\tau_R$ for $\Psi_0(t)=J_0(t/\tau_R)$. For $\tau/\tau_R=0.1,1,10$, a straight line was observed (see Fig.~\ref{fig:1}).}
\label{table:2}
\end{table}

\begin{table}[h!]
\centering
\begin{tabular}{|c|c|c|c|}
\hline
\multicolumn{4}{|c|}{$\Psi_0(t)=\text{sinc}(t/\tau_R)$}  \\
\hline
$\tau/\tau_R$ & $0.1$ & $1$ & $10$ \\
\hline
$W_{\rm irr}^*$ &0.4999$\pm$0.0001  &0.4864$\pm$0.0001  &0.14748$\pm$0.00001  \\
\hline
$W_{\mathrm{exact}}^*$ &0.499861  &0.486385  &0.147444  \\
\hline
QC &$10^{-15}$  & $10^{-9}$  &$10^{-15}$  \\
\hline
\end{tabular}
\caption{Values of $W_{\rm irr}^*$, quadratic constraint (QC), and $W_{\mathrm{exact}}^*$ for different $\tau/\tau_R$ for $\Psi_0(t)=\text{sinc}(t/\tau_R)$. For $\tau/\tau_R=0.1,1,10$, a straight line was observed (see Fig.~\ref{fig:1}).}
\label{table:3}
\end{table}

\begin{table}[h!]
\centering
\begin{tabular}{|c|c|c|c|}
\hline
\multicolumn{4}{|c|}{$\Psi_0(t)=\cos{(\omega t)}$}  \\
\hline
$\omega\tau$ & $0.1$ & $1$ & $10$ \\
\hline
$W_{\rm irr}^*$ &0.4996$\pm$0.0001  &0.4597$\pm$0.0001  &0.0183864$\pm$0.000002  \\
\hline
$W_{\mathrm{exact}}^*$ &0.499583  &0.459698  &0.0183907  \\
\hline
$\text{QC}$ &$10^{-9}$  & $10^{-8}$  &$10^{-6}$  \\
\hline
\end{tabular}
\caption{Values of $W_{\rm irr}^*$, quadratic constraint (QC), and $W_{\mathrm{exact}}^*$ for different $\omega\tau$ for $\Psi_0(t)=\cos{(\omega t)}$. For $\omega\tau=0.1,1,10$, a straight line was observed (see Fig.~\ref{fig:1}).}
\label{table:4}
\end{table}

\section{Final remarks}

In this work, we investigated the problem of minimizing irreversible work in weakly driven systems within linear response, under constraints that ensure physical admissibility of the protocol. By fixing both the total variation and imposing a quadratic bound on the driving speed, the optimization problem avoids the singular behaviors that often arise in unconstrained formulations.

A central aspect of the analysis concerns the structure of the relaxation function. Rather than viewing the periodic representation as an external assumption, we have clarified that it follows directly from two intrinsic properties of the kernel. First, the dependence on time differences implies that the relevant domain is naturally the symmetric interval from $[-\tau,\tau]$. Second, the evenness of the relaxation function ensures continuity at the boundaries of this interval. Together, these features define the natural space on which the operator acts.

Within this setting, the periodic extension appears as the minimal closure that preserves these properties while restoring full translational invariance. As a consequence, the integral operator becomes a convolution operator on a periodic domain, and the Fourier basis emerges as its natural spectral representation. In this sense, the diagonalization of the problem is not a technical convenience, but a direct manifestation of the underlying structure of the kernel.

This perspective allows the optimization problem to be reduced to a shifted eigenvalue equation whose solution is determined by the spectral properties of the relaxation function. The positivity of the kernel selects the zero mode as the global minimizer under the imposed constraints, leading to a constant driving speed and therefore to a linear protocol. While linear protocols are known to arise in linear-response frameworks under conditions such as constant thermodynamic metric or slow driving~\cite{Sivak2012}, in those cases linearity follows from the locality of the functional or geometric arguments. In contrast, here linearity emerges as a consequence of the spectral structure of a nonlocal quadratic functional, where the positivity of the relaxation kernel selects the zero mode as the global minimizer.

The corresponding irreversible work takes a simple and universal form, depending only on the integrated relaxation function. This shows that, within linear response, dissipation is governed by a global measure of memory rather than by the detailed temporal shape of the correlations. This suggests that similar spectral mechanisms may underlie optimal protocols in more general nonequilibrium settings, where memory effects and constraints compete to shape the structure of dissipation. Also, the scaling with the protocol duration is consistent with the quasistatic limit, ensuring vanishing dissipation for slow processes. 

The numerical analysis supports these conclusions across a wide class of relaxation functions, including oscillatory and decaying kernels. In all cases considered, the optimal protocol converges to a linear form, and the computed work agrees with the analytical prediction with high precision. This indicates that the result is robust and does not depend on specific features of the kernel.

Overall, the results establish that the emergence of linear protocols is a direct consequence of the structural properties of the relaxation function. The role of periodicity is not to impose a mathematical framework, but to reveal the natural domain and symmetry already encoded in the problem. From an operational perspective, this periodic representation reflects how irreversible work is defined in practice, namely as an ensemble average over many realizations of the same finite-time protocol, each performed over the interval $[0,\tau]$ and followed by a reset of the system. This repeated structure introduces an effective block decomposition in time, where no realization is distinguished from another, and time differences are naturally represented on a periodic domain.

In this sense, periodicity emerges as a consistent representation of the statistical structure of repeated measurements, rather than a property of the underlying dynamics. This perspective provides a clearer conceptual foundation for optimal control in weakly driven systems and suggests a natural path for extending the analysis beyond linear response, where the interplay between memory, symmetry, and constraints may lead to richer behaviors.

\section*{Data availability}

The code used in this work is available at Ref.~\cite{pnaze_bcwd}.

\bibliographystyle{apsrev4-2}
\bibliography{BCWD}

\onecolumngrid

\appendix

\section{Periodic extension of the Euler-Lagrange equation}
\label{app:A}

Importantly, the periodic extension does not modify the variational problem on the original interval. Let $\mathcal{K}$ denote the operator defined on $[0,\tau]$ by
\be
(\mathcal{K}v)(t) = \int_0^\tau \Psi_0(t-u)\,v(u)\,du,
\ee
and let $\tilde{\mathcal{K}}$ denote the operator obtained by periodically extending the kernel and extending $v(t)$ evenly to $[-\tau,\tau]$. By construction, for any admissible function $v$ and for all $t \in [0,\tau]$, one has
\be
(\tilde{\mathcal{K}}v)(t) = (\mathcal{K}v)(t).
\ee
Therefore, the extended Euler–Lagrange equation, when restricted to $[0,\tau]$, is exactly equivalent to the original one. Since the kernel is positive definite, the minimizer is unique, and the solution in the extended setting necessarily coincides with that of the original problem.

\section{Representation in a Chebyshev basis}
\label{app:B}

In this appendix, we briefly examine the representation of the variational problem in a basis that is natural on a finite interval, namely, Chebyshev polynomials. This provides a useful contrast with the Fourier basis and clarifies the role of translational invariance in the diagonalization of the operator.

We map the interval $t \in [0,\tau]$ onto $x \in [-1,1]$ through
\be
x = \frac{2t}{\tau} - 1,
\ee
and expand the protocol derivative as
\be
v(t) \equiv v(x) = \sum_{n=0}^\infty d_n\, T_n(x),
\ee
where $T_n(x)$ are Chebyshev polynomials of the first kind.

The irreversible work functional becomes
\be
W_{\mathrm{irr}} = \frac{1}{2} \int_{-1}^1 \int_{-1}^1 \tilde{\Psi}(x,y)\, v(x)\, v(y)\, J(x)\, J(y)\, dx\, dy,
\ee
where $\tilde{\Psi}(x,y) = \Psi_0\!\big(t(x)-t(y)\big)$ and $J(x)=\tfrac{\tau}{2}$ is the Jacobian of the transformation.

Substituting the expansion, one obtains the quadratic form
\be
W_{\mathrm{irr}} = \frac{1}{2} \sum_{m,n=0}^\infty d_m\, d_n\, K_{mn},
\ee
with matrix elements
\be
K_{mn} = \int_{-1}^1 \int_{-1}^1 \tilde{\Psi}(x,y)\, T_m(x)\, T_n(y)\, J(x)\, J(y)\, dx\, dy.
\ee

In contrast to the Fourier representation discussed in the main text, the matrix $K_{mn}$ is not diagonal. This reflects the fact that Chebyshev polynomials are not eigenfunctions of convolution operators, even when the relaxation function is extended periodically. As a result, the shifted eigenvalue problem derived from the Euler-Lagrange equation takes the form of a full matrix equation,
\be
\sum_{n=0}^\infty K_{mn} d_n = \alpha_1 M_m + \alpha_2 d_m,
\ee
where
\be
M_m = \int_{-1}^1 T_m(x)\, J(x)\, dx,
\ee
arises from the linear constraint.

Despite this non-diagonal structure, the operator defined by the kernel remains symmetric and positive definite, ensuring that its smallest eigenvalue determines the global minimum of the functional. The corresponding eigenfunction is the constant function, which is exactly represented by the $T_0(x)$ mode. Therefore, the optimal solution remains
\be
v^*(t) = \frac{\delta \lambda}{\tau},
\ee
in agreement with the Fourier-based analysis.

This comparison shows that the simplicity of the solution in the main text is not a consequence of the choice of basis, but rather of the translational invariance restored by the periodic representation of the relaxation function. In the Fourier basis, this structure leads to diagonalization, making the optimal solution immediately transparent, whereas in other bases, such as the Chebyshev basis, the same result follows from the spectral properties of a full matrix representation.

\end{document}